\documentclass{PoS}

\usepackage{amsmath}
\usepackage{amsthm}
\usepackage{amsfonts}
\usepackage{dsfont}
\usepackage{graphics}
\usepackage{graphicx}
\usepackage{epsfig}
\usepackage{amssymb}
\usepackage{wrapfig}

\title{Prediction of positive parity $B_s$ mesons and search for the $X(5568)$}

\ShortTitle{Positive parity $B_s$ mesons and the $X(5568)$}

\author{\speaker{Daniel Mohler}\\
        Helmholtz-Institut Mainz, 55099 Mainz, Germany\\
        Johannes Gutenberg Universit\"at Mainz, 55099 Mainz, Germany\\
        E-mail: \email{mohler@kph.uni-mainz.de}}

\author{C.~B.~Lang\\
        Institute of Physics,  University of Graz, A--8010 Graz, Austria\\
        E-mail: \email{christian.lang@uni-graz.at}}

\author{Sasa Prelovsek\\
        Department of Physics, University of Ljubljana, 1000 Ljubljana, Slovenia\\
        Jozef Stefan Institute, 1000 Ljubljana, Slovenia\\
        E-mail: \email{sasa.prelovsek@ijs.si}}

\abstract{We use a combination of quark-antiquark and $B^{(*)}K$ interpolating
  fields to predict the mass of two QCD bound states below the $B^*K$
  threshold in the quantum channels $J^P=0^+$ and $1^+$. The mesons correspond
  to the b-quark cousins of the $D_{s0}^*(2317)$ and $D_{s1}(2460)$ and have
  not yet been observed  in experiment, even though they are expected to be
  found by LHCb. In addition to these predictions, we obtain excellent agreement of the remaining p-wave energy levels with the known $B_{s1}(5830)$ and $B_{s2}^*(5840)$ mesons. The results from our first principles calculation are compared to previous model-based estimates. More recently the D0 collaboration claimed the existence of an exotic resonance $X(5568)$ with exotic flavor content $\bar{b}s\bar{d}u$. If such a state with $J^P=0^+$ exists, only the decay into $B_s\pi$ is open which makes a lattice search for this state much cleaner and simpler than for other exotic candidates involving heavy quarks. We conclude, however, that we do not find such a candidate in agreement with a recent LHCb result.}

\FullConference{34th annual International Symposium on Lattice Field Theory\\
		24-30 July 2016\\
		University of Southampton, UK}

\begin{document}

In these proceedings we summarize two recently published lattice QCD studies
\cite{Lang:2015hza,Lang:2016jpk} of states close to multi-particle thresholds.

\section{Prediction of positive parity $B_s$ mesons}

The discovery of the $D_{s0}^*(2317)$ by BaBar \cite{Aubert:2003fg} and the subsequent
discovery of the $D_{s1}(2460)$ more than 10 years ago revealed an unexpected
peculiarity: unlike expected by potential models, these states turned out to
be narrow states below the $DK$ and $D^*K$ thresholds. Moreover their mass is
roughly equal to the mass of their non-strange cousins, which immediately
sparked speculations about their structure in terms of quark content, with
popular options including both tetraquark and molecular structures.

The corresponding $J^P=0^+$ and $1^+$ states in the spectrum of $B_s$ hadrons have not been
established in experiment. Given the success of recent lattice QCD
calculations of the  $D_{s0}^*(2317)$ and $D_{s1}(2460)$ \cite{Mohler:2013rwa,Lang:2014yfa}, it is
therefore interesting to see if a prediction of these positive parity $B_s$
states from lattice QCD is feasible. 

\begin{table}[bh]
\begin{center}
\begin{tabular}{ccccccc}
$N_L^3\times N_T$ & $N_f$ & $a$[fm] & $L$[fm] & \#configs & $m_\pi$[MeV] & $m_K$[MeV]\\ 
\hline
$32^3\times64$ & 2+1 & 0.0907(13) & 2.90 & 196 & 156(7)(2) & 504(1)(7)\\
\end{tabular}
\caption{Gauge configurations used for the simulations in these proceedings.}
\label{configs}
\end{center}
\end{table}

\subsection{Lattice techniques}

For this study we use the 2+1 flavor gauge configurations with Wilson-Clover quarks generated by the
PACS-CS collaboration \cite{Aoki:2008sm}. Table \ref{configs} shows details of the ensemble used
in our simulation. Our quark sources are smeared with a Gaussian-like envelope
as produced by use of the stochastic distillation technique
\cite{Morningstar:2011ka}. For the heavy b-quarks in the Fermilab
interpretation \cite{ElKhadra:1996mp}, we tune the heavy-quark hopping parameter
  $\kappa_b$ for the spin averaged kinetic mass
$M_{\overline{B_s}}=(M_{B_s}+3M_{B_s^*})/4$ to assume its physical
value. The energy splittings we determine are expected to be close to physical
in this setup. For technical details on the tuning of the
heavy-quark hopping parameter please refer to \cite{Lang:2014yfa,Lang:2015hza}. We work with a
partially quenched strange quark and used the $\phi$ meson and $\eta_{s}$ to
set the strange quark mass, obtaining $\kappa_s=0.13666$ \cite{Lang:2014yfa}. Table
\ref{splittings} shows examples of mass splittings extracted with this setup. Notice
that the uncertainties provided in this table are statistical and scale-setting
uncertainties only. Nevertheless the agreement with experiment is mostly
excellent, indicating that the remaining discretization
effects are small.

\begin{table}[tbh]
\begin{center}
\begin{tabular}{|c|c|c|}
\hline
                       &       Lattice [MeV]     & Exp. [MeV]       \cr
\hline
$m_{B^*}-m_B$ & 46.8(7.0)(0.7) & 45.78(35)\cr
$m_{B_{s^*}}-m_{B_s}$ & 47.1(1.5)(0.7)& $48.7^{+2.3}_{-2.1}$\cr
$m_{B_s}-m_{B}$ & 81.5(4.1)(1.2) & 87.35(23)\cr
$m_Y-m_{\eta_b}$ &  44.2(0.3)(0.6) & 62.3(3.2)\cr
$2m_{\overline{B}}-m_{\overline{\bar{b}b}}$ & 1190(11)(17) & 1182.7(1.0)\cr
$2m_{\overline{B_s}}-m_{\overline{\bar{b}b}}$ & 1353(2)(19)& 1361.7(3.4)\cr
$2m_{B_c}-m_{\eta_b}-m_{\eta_c}$ & 169.4(0.4)(2.4) & 167.3(4.9) \cr
\hline
\end{tabular}
\caption{Selected mass splittings (in MeV) of mesons involving bottom quarks compared to the values from the PDG \cite{Agashe:2014kda}. A bar denotes spin average. Errors are statistical and scale-setting only.\label{splittings}}
\end{center}
\end{table}

\begin{figure}[tb]
\begin{center}
\includegraphics[width=0.47\textwidth,clip]{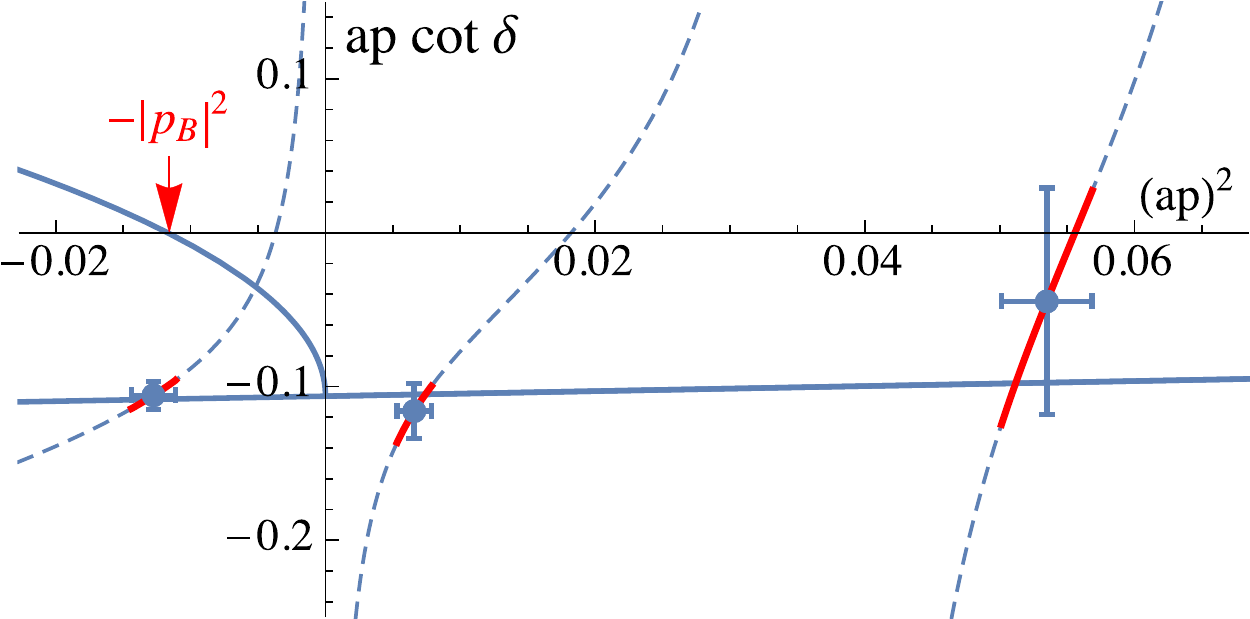}
\includegraphics[width=0.47\textwidth,clip]{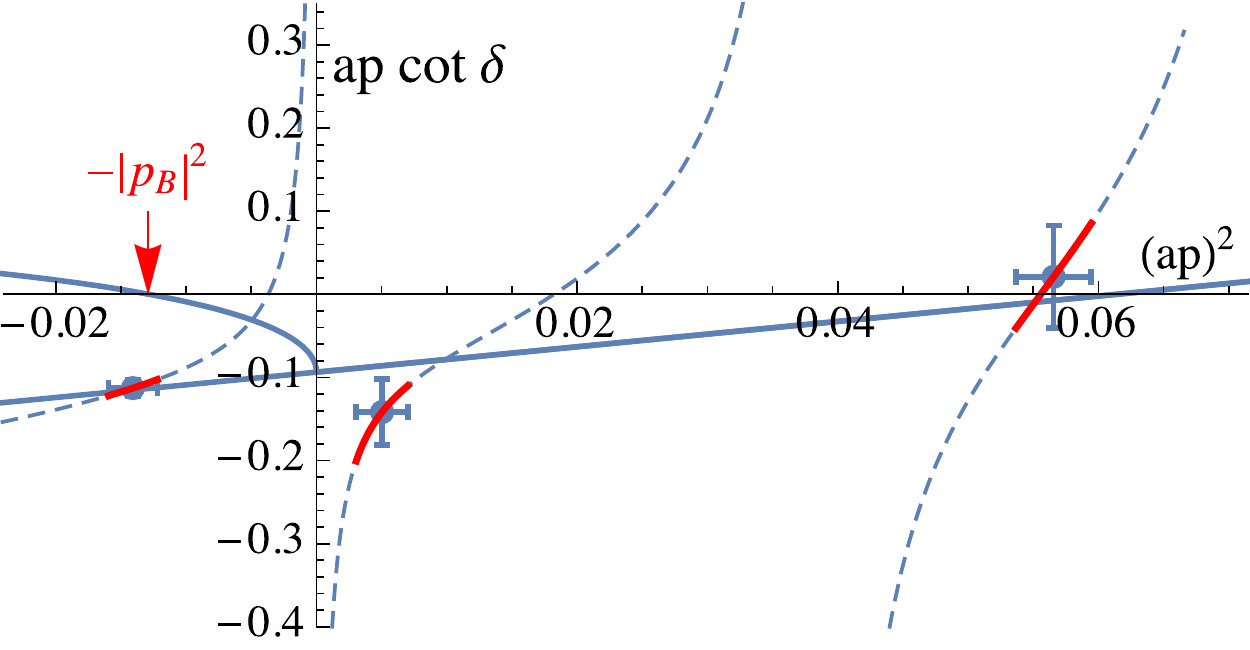}
\caption{ Plots of $ap \cot\delta(p)$ vs. $(ap)^2$ for $B^{(*)}K$ scattering in $s$-wave. Circles are  values from our simulation; red lines indicate the error band following
the L\"uscher curves (broken lines). The full line gives the linear fit to the points. Below threshold $|p|$ is added and the zero of the
combination indicates the bound state position in infinite volume. Displayed
uncertainties are statistical only. \label{fig:effrange}}
\end{center}
\end{figure}

For the construction of the correlation matrix used to extract the finite
volume energies, our study takes into account both quark-antiquark as well as
B-K structures. The basis is similar to our study of the $D_s$ spectrum
\cite{Mohler:2013rwa,Lang:2014yfa}, where this approach allowed us to obtain
reliable energy levels for the $D_{s0}^{*}(2317)$ and $D_{s1}(2460)$. For
elastic s-wave scattering the L\"uscher relation \cite{Luscher} relating the finite volume
spectrum to the phase shift $\delta$ of the infinite volume scattering amplitude is
given by 
\begin{align}
%K^{-1}=
p \cot
\delta(p)&=\frac{2}{\sqrt{\pi}L}Z_{00}(1;q^2)\approx\frac{1}{a_0}+\frac{1}{2}r_0p^2\;.
\label{eq:luescher_z}
\end{align}
We perform an effective range approximation with the s-wave scattering
length $a_0$ and effective range $r_0$. The resulting parameters and the mass
of the resulting binding momentum (from $\cot(\delta(p))=i$) are shown in Figure
\ref{fig:effrange}. We obtain
\begin{align}
a_0^{BK}&=-0.85(10)\,\mathrm{fm} &a_0^{B^*K}=-0.97(16) \,\mathrm{fm}\;\,\\
r_0^{BK}&=0.03(15)  \,\mathrm{fm} &r_0^{B^*K}=0.28(15)  \,\mathrm{fm}\;\,\nonumber\\
M_{B_{s0}^*}&=5.711(13)\,\mathrm{GeV} &M_{B_{s1}}=5.750(17)  \,\mathrm{GeV}\;\,\nonumber
\end{align}
where the uncertainty on the bound state mass is statistical only. A full uncertainty
estimate is given in Table \ref{errors} and explained in more detail in \cite{Lang:2015hza}.

\begin{table}[tbp]
\begin{center}
\begin{tabular}{cc}
source of uncertainty &  expected size [MeV]\\
\hline
heavy-quark discretization & 12\\
finite volume effects & 8\\
unphysical Kaon, isospin \& EM & 11\\
b-quark tuning & 3\\
dispersion relation & 2\\
spin-average (experiment) & 2\\
scale uncertainty & 1\\
3 pt vs. 2 pt linear fit & 2\\
\hline
total (added in quadrature) & 19\\
\end{tabular}
\end{center}
\caption{Systematic uncertainties in the mass determination of the
  below-threshold states with quantum numbers $J^P=0^+, 1^+$. The heavy-quark
  discretization effects are quantified by calculating the Fermilab-method
  mass mismatches and employing HQET power counting \cite{Oktay:2008ex} with
  $\Lambda=700$~MeV. The finite volume uncertainties are estimated
  conservatively by the difference of the lowest energy level and the complex pole position. The last line gives the effect of using only the two points near threshold for the effective range fit. The total uncertainty has been obtained by adding the single contributions in quadrature.\label{errors}}
\end{table}

\begin{figure}[tbp]
\begin{center}
\includegraphics[width=0.6\textwidth,clip]{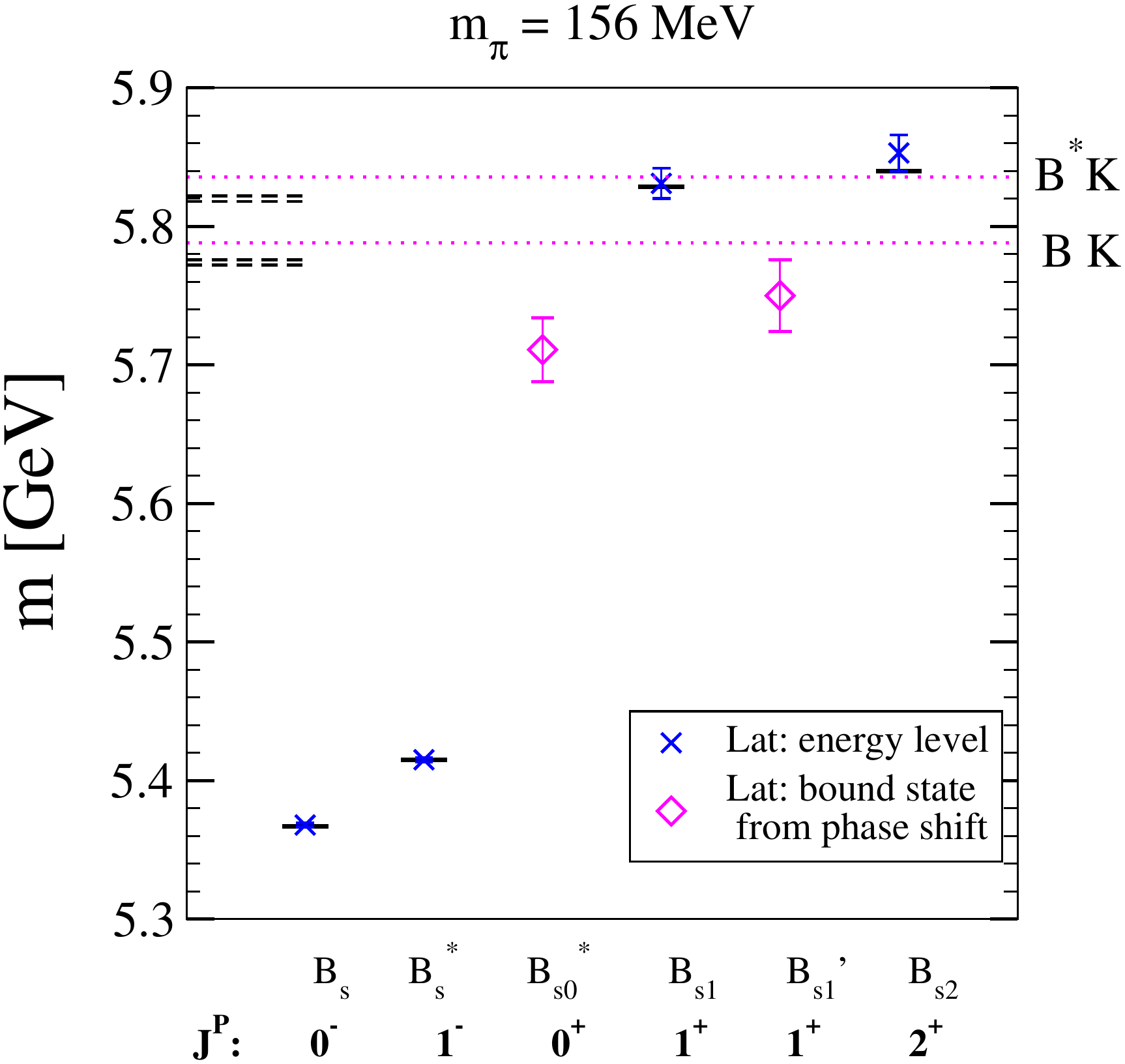}
\end{center}
\caption{ Spectrum of s-wave and p-wave $B_s$ states from our simulation. The
  blue states are naive energy levels, while the bound state energy of the
  states in magenta results from an effective range approximation of the phase
shift data close to threshold. The black lines are the energy levels from the
PDG \cite{Agashe:2014kda}. The error bars on the blue states are
statistical only, while the errors on the magenta states show the full
(statistical plus systematic) uncertainties.}
\label{fig:finalbs}
\end{figure}

\subsection{Resulting prediction of positive parity $B_s$ mesons}

Figure \ref{fig:finalbs} shows our final results for the spectrum of s-wave and p-wave
$B_s$ states. For values of masses in MeV we quote $M=\Delta
M^{\mathrm{lat}}+M_{\overline{B_s}}^{\mathrm{exp}}$ where we substitute the experimental $B_s$ spin average in
accordance with our tuning. The states with blue symbols result from a naive determination
of the finite volume energy levels (statistical uncertainty only). Notice that
the $j={\frac{3}{2}}$ states agree well with the experimental $B_{s1}(5830)$
and $B_{s2}^*(5840)$ as determined by CDF/D0 and LHCb \cite{Agashe:2014kda}. The $B_s$ states
with magenta symbols indicate the bound state positions extracted using
L\"uscher's method and taking into account the sources of uncertainty detailed
in Table \ref{errors}. Notice that our Lattice QCD calculations yields bound
states well below the $B^{(*)}K$ thresholds.

\section{$B_s \pi^+$ scattering and search for the X(5568)}

\begin{figure}[tbp]
\begin{center}
\includegraphics[width=0.43\textwidth,clip]{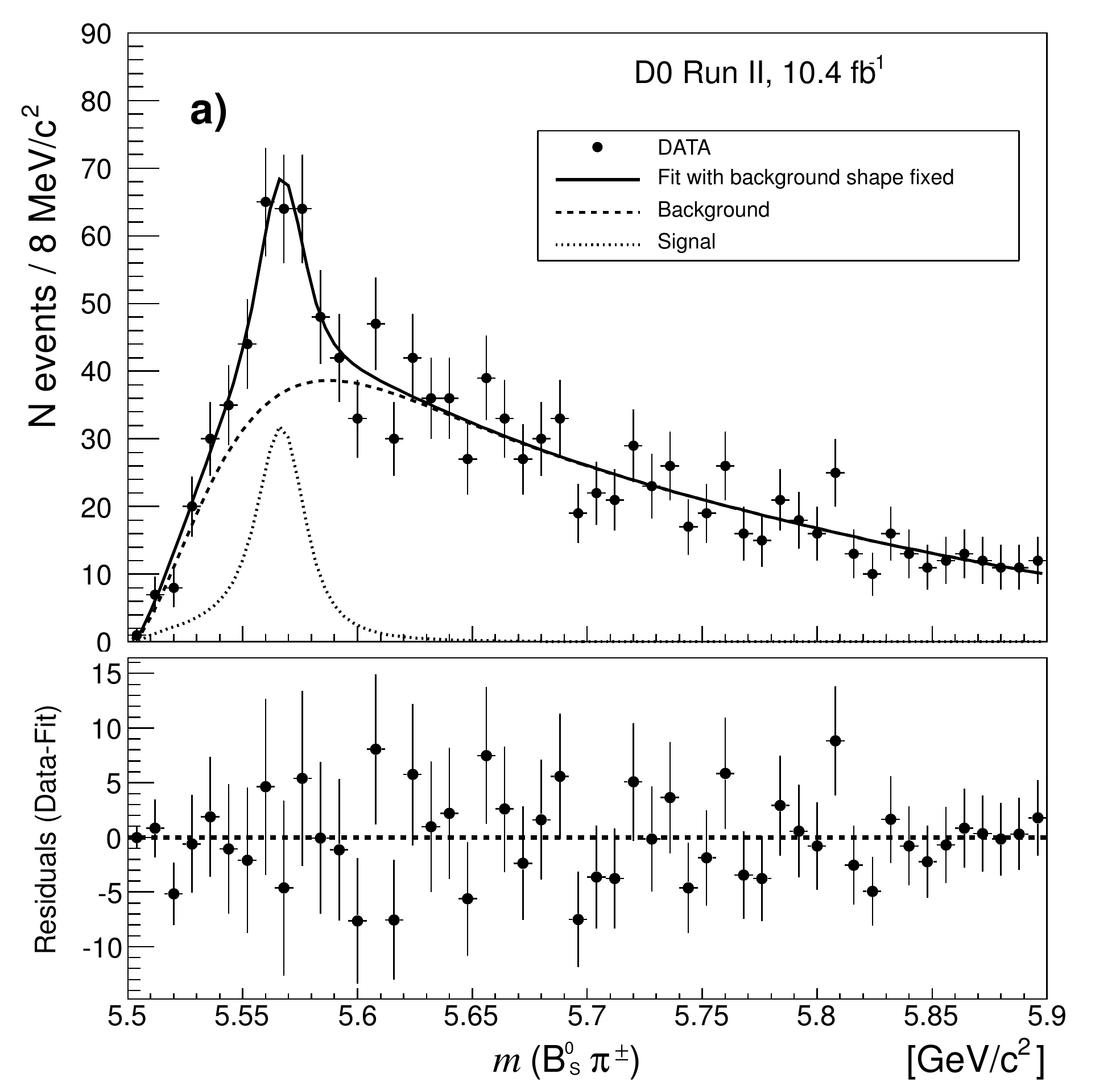}
\includegraphics[width=0.53\textwidth,clip]{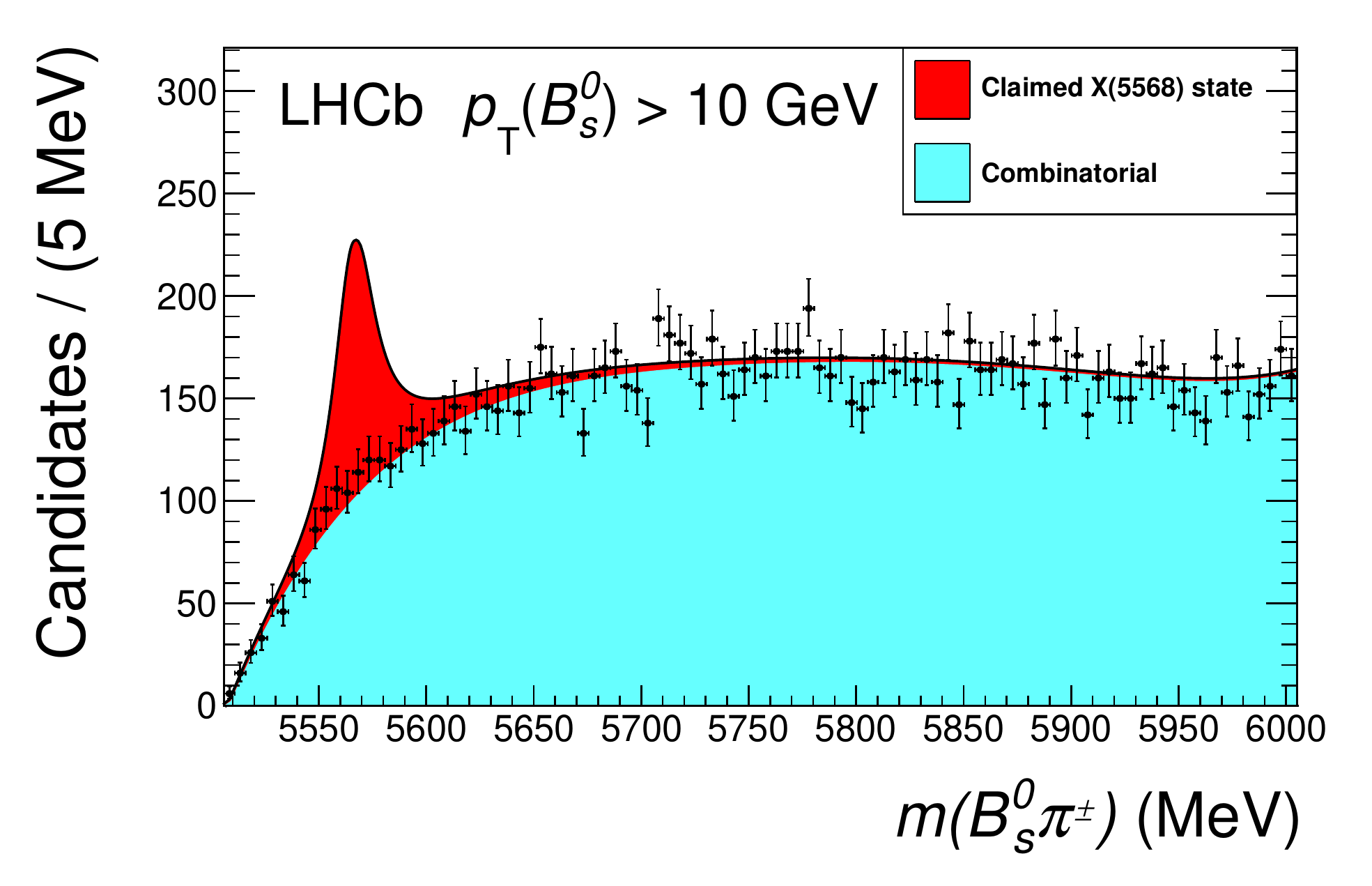}
\end{center}
\caption{Left pane: $B_s^0\pi^\pm$ invariant mass distribution from D0 \cite{D0:2016mwd} (after
  applying a cone cut). Right pane: $B_s^0\pi^\pm$ invariant mass distribution
  by LHCb \cite{Aaij:2016iev} shown in black symbols with a signal
  component corresponding to $\rho_x=8.6\%$ as observed by D0 shown in red.}
\label{fig:x5568_exp}
\end{figure}

Recently, the D0 collaboration has reported evidence for a peak in the $B_s\pi^+$
  invariant mass not far above threshold \cite{D0:2016mwd}. This peak is
  attributed to a resonance dubbed X(5568)with the resonance mass $m_X$ and
  width $\Gamma_x$,
\begin{align}
m_X&=5567.8\pm2.9^{+0.9}_{-1.9}\;\mathrm{MeV}\;,\\
\Gamma_X&=21.9\pm6.4^{+5.0}_{-2.5}\;\mathrm{MeV}\;.\nonumber
\end{align}

\begin{wrapfigure}{r}{0.5\textwidth}
\includegraphics[width=0.45\textwidth,clip]{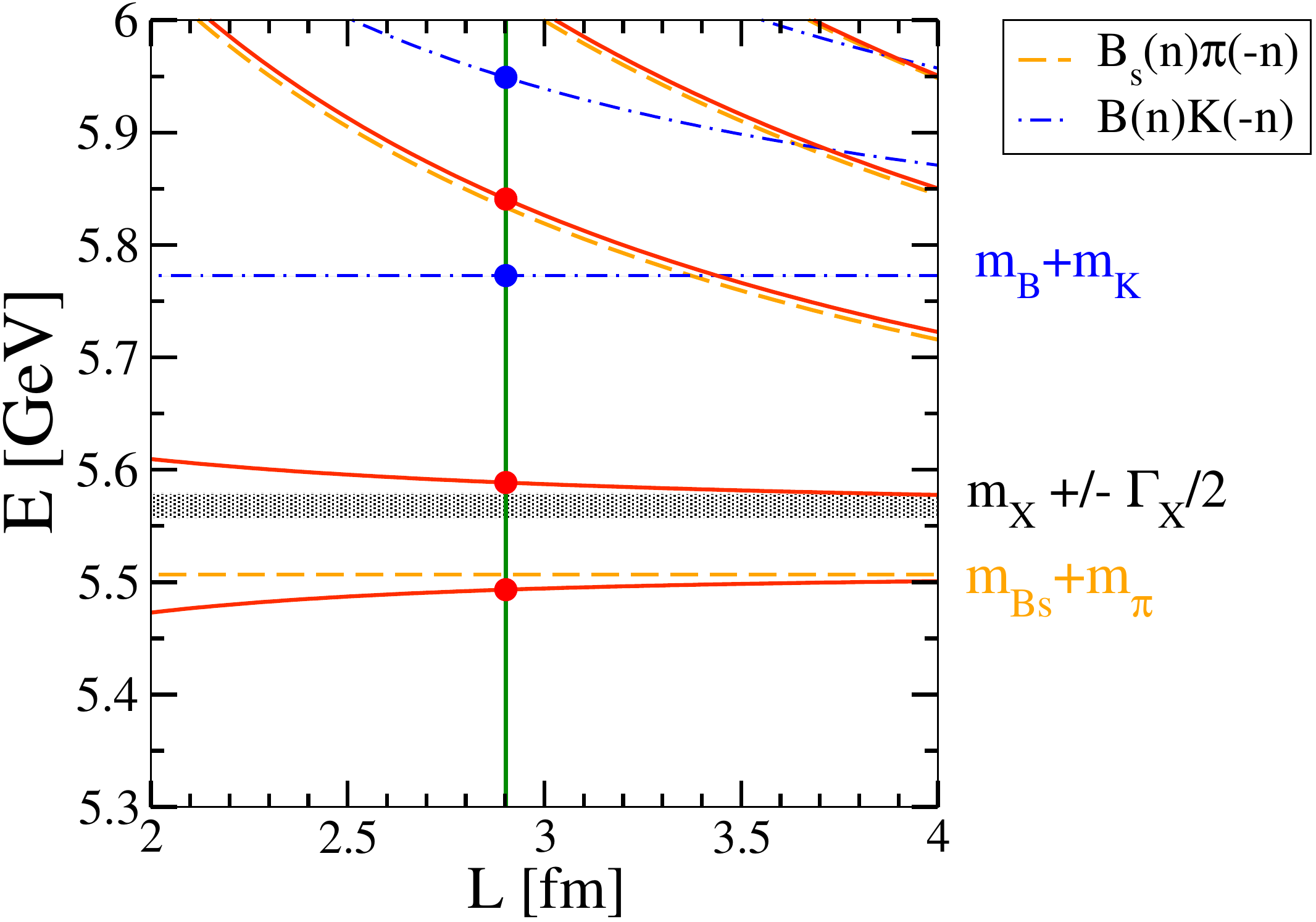}
\caption{Analytic predictions for energies $E(L)$ of  eigenstates as a
  function of lattice size $L$.}
\label{fig:analytic}
\end{wrapfigure}

Decay of this resonance into $B_s\pi^+$ implies an exotic flavor structure
with the minimal quark content $\bar b s \bar d u$. Most model studies which
accommodate a X(5568) propose spin-parity quantum numbers $J^P=0^+$. Short
after D0 reported their results, the LHCb collaboration investigated the cross-section
as a function of  the $B_s\pi^+$ invariant mass with increased statistics and did 
not find any peak in the  same region \cite{Aaij:2016iev}. Figure
\ref{fig:x5568_exp} shows both the plot from D0 (left pane) and the data from
LHCb (right pane), where the red shaded region illustrates the signal
expectation given the ratio of yields $\rho_x$ determined by D0.

\subsection{Expected signature for a resonance in $B_s\pi^+$}

The presence of an elastic resonance with the parameters of the $X(5568)$
would lead to a characteristic pattern of finite volume energy levels
corresponding to QCD eigenstates with given quantum numbers for finite spatial size $L$.

\begin{wrapfigure}{r}{0.45\textwidth}
\includegraphics[width=0.45\textwidth,clip]{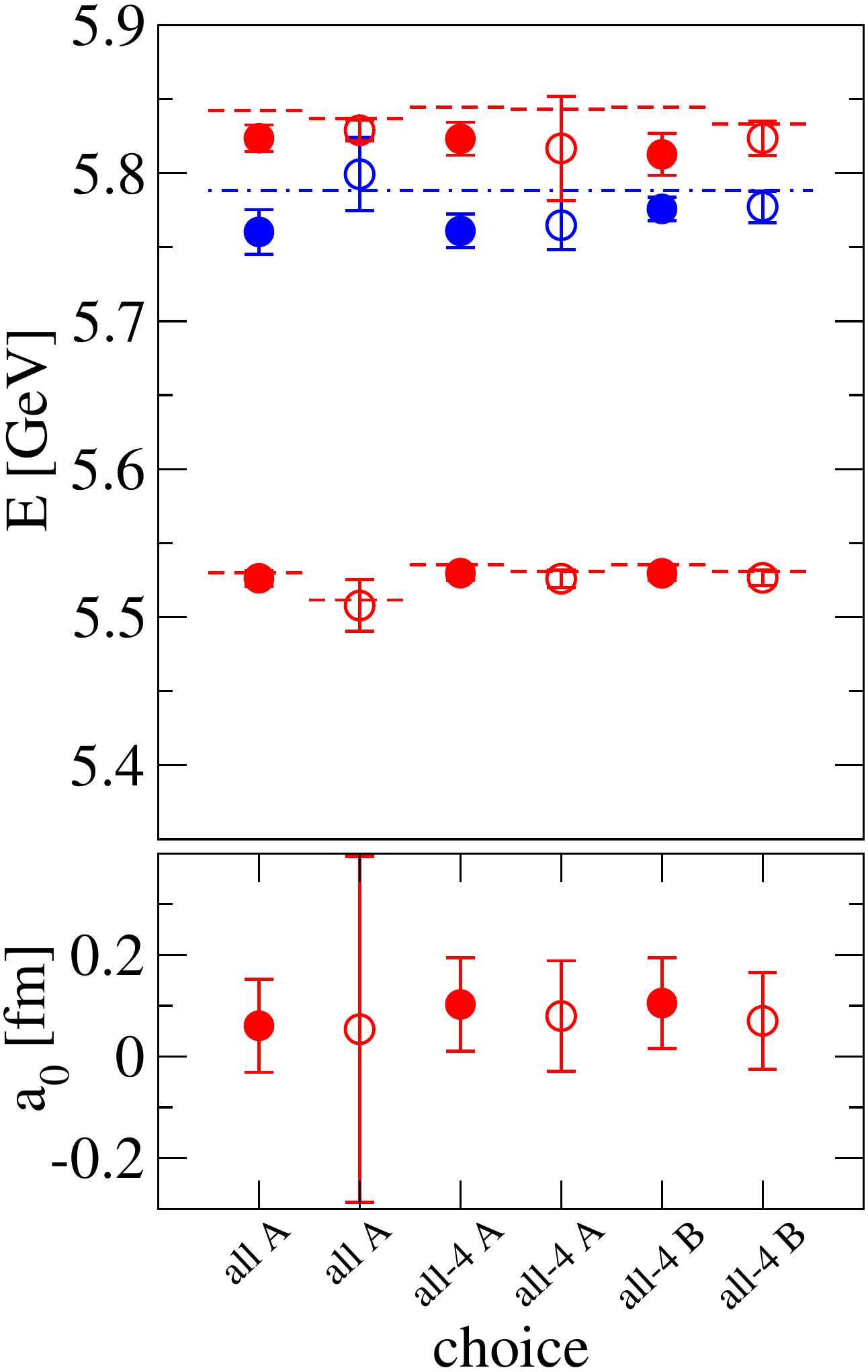}
\caption{The eigenenergies of the $\bar bs\bar du$ system with $J^P=0^+$
  from a lattice simulation for the choices detailed in the text.}
\label{fig:lattice}
\end{wrapfigure}

Figure \ref{fig:analytic} shows analytic predictions for energies of
eigenstates for an elastic resonance in $B_s\pi$ (with $J^P=0^+$) as a
function of the lattice size $L$ as determined from L\"uscher's formalism
\cite{Luscher}. Red solid lines are $B_s\pi$ eigenstates in
  the scenario with resonance  $X(5568)$ ; orange dashed
  lines are $B_s\pi$  eigenstates when $B_s$ and $\pi$ do not interact;
  blue dot-dashed lines are $B^+\bar K^0$ eigenstates  when $B^+$ and $\bar
  K^0$ do not interact; the grey band indicates the position of $X(5568)$ from
  the D0 experiment  \cite{D0:2016mwd}. The lattice size $L=2.9~$fm, used in
  our simulation, is marked by the vertical line. Note that the resonant
  scenario  predicts an eigenstate near $E\simeq m_X$ (red solid), while there is no such eigenstate for $L=2-4~$fm in
a scenario with no or small interaction between $B_s$ and $\pi^+$ (orange
dashed). In the unlikely scenario of a deeply  bound $BK$ state, the
simulation would result in an eigenstate with  $E\approx m_X$ up to exponentially small corrections in $L$.

\subsection{Simulation details}

In our simulation we use the PACS-CS ensemble \cite{Aoki:2008sm} from Table \ref{configs}. The interpolator basis 

 \begin{align}
O_{1,2}^{B_s(0)\pi(0)}&=\left[\,\bar{b}\Gamma_{1,2} s\,\right](\mathbf{p}=0)\left[\,\bar{d}\Gamma_{1,2} u\,\right](\mathbf{p}=0)\;,\nonumber\\ 
O_{1,2}^{B_s(1)\pi(-1)}&=\!\!\!\!\!\!\!\!\!\sum_{\mathbf{p}=\pm\mathbf{e_{x,y,z}}~2\pi/L}\!\!\!\!\!\!\! \left[\bar{b}\Gamma_{1,2} s\right](\mathbf{p})\left[\bar{d}\Gamma_{1,2} u\right](-\mathbf{p})\nonumber\;,\\ 
O_{1,2}^{B(0)K(0)}&=\left[\,\bar{b}\Gamma_{1,2} u\,\right](\mathbf{p}=0)\left[\,\bar{d}\Gamma_{1,2} s\,\right](\mathbf{p}=0)\nonumber\;,
\end{align}
consisting of both $B_s\pi$ and $BK$ interpolators, is employed.

Figure \ref{fig:lattice} shows the eigenstates determined from our simulation
for various choices. The sets with full symbols are from correlated fits
while open symbols result from uncorrelated fits. Notation ``all'' refers to
the full set of gauge configurations while ``all-4'' refers to the set with
four (close to exceptional) gauge configurations removed. Set A is from 
interpolator basis
$O_{1}^{B_s(0)\pi(0)}$, $O_{1}^{B_s(1)\pi(-1)}$, $O_{1}^{B(0)K(0)}$ while set B
results from a larger basis
$O_{1}^{B_s(0)\pi(0)},O_{1,2}^{B_s(1)\pi(-1)},O_{1,2}^{B(0)K(0)}$. All choices
consistently result in a small scattering length $a_0$ consistent with 0
within error.

\subsection{Concludions from comparing analytic predictions and lattice energy levels}

\begin{wrapfigure}{r}{0.5\textwidth}
\includegraphics[width=0.5\textwidth,clip]{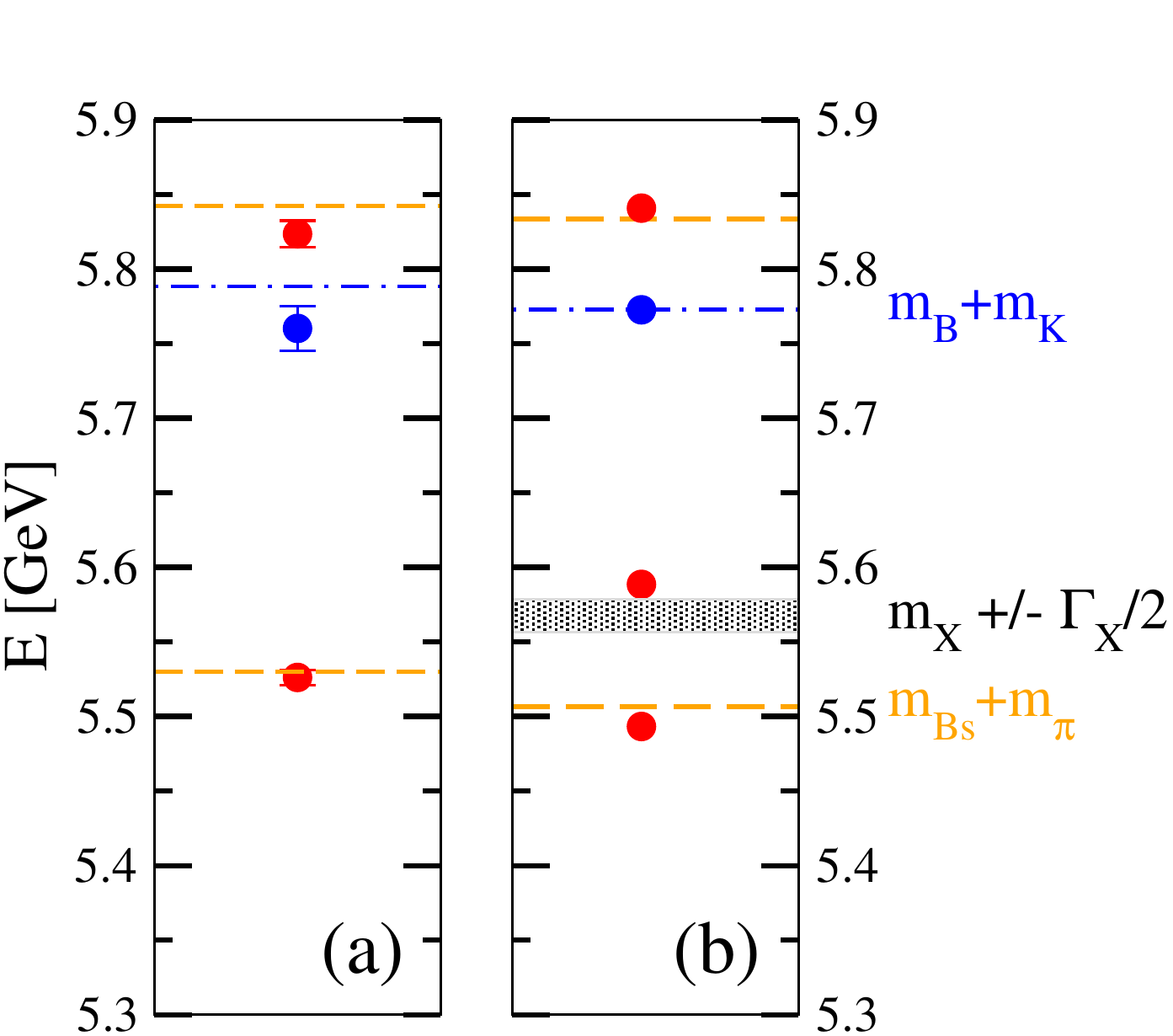}
\caption{(a) The eigenenergies of  the $\bar bs\bar du$ system with $J^P=0^+$
  from our lattice simulation    and  (b) an  analytic prediction based on
  $X(5568)$, both at lattice size $L=2.9~$fm.  The horizontal lines show energies of eigenstates $B_s(0)\pi^+(0)$, $B^+(0)\bar K^0(0)$ and $B_s(1)\pi^+(-1)$ in absence of interactions; momenta in units of $2\pi/L$ are given in parenthesis.  
The pane (a) shows the energies
$E=E^{lat}_n-E_{\overline{B_s}}^{lat}+E^{exp}_{\overline{B_s}}$ with the
spin-averaged $B_s$ ground state set to its experiment value.  The pane (b) is based on the
experimental mass of the $X(5568)$ \cite{D0:2016mwd}, given by the grey band,
and experimental masses of other particles.}
\label{fig:combined}
\end{wrapfigure}

Figure \ref{fig:combined} shows the eigenenergies of the $\bar bs\bar du$ system with
$J^P=0^+$ calculated on the lattice (left pane) compared to the analytic
prediction based on the $X(5568)$ as observed  by D0 (right pane).
Unlike expected for the case of a resonance with the
parameters of the X(5568), our lattice simulation at close-to-physical quark masses does not yield a
  second low-lying energy level. Our results therefore do not support the existence of $X(5568)$ with
  $J^P=0^+$. Instead, the results appear closer to the limit where $B_s$ and
  $\pi$ do not interact significantly, leading to a $B_s\pi$ scattering length
  compatible with 0 within errors.

\begin{thebibliography}{99}
\bibitem{Lang:2015hza} 
  C.~B.~Lang, D.~Mohler, S.~Prelovsek and R.~M.~Woloshyn,
  %``Predicting positive parity B$_s$ mesons from lattice QCD,''
  Phys.\ Lett.\ B {\bf 750}, 17 (2015)
  doi:10.1016/j.physletb.2015.08.038.
  %%CITATION = doi:10.1016/j.physletb.2015.08.038;%%
  %18 citations counted in INSPIRE as of 24 Oct 2016
\bibitem{Lang:2016jpk} 
  C.~B.~Lang, D.~Mohler and S.~Prelovsek,
  %``$B_s\pi^+$ scattering and search for X(5568) with lattice QCD,''
  Phys.\ Rev.\ D {\bf 94}, 074509 (2016)
  doi:10.1103/PhysRevD.94.074509.
  %%CITATION = doi:10.1103/PhysRevD.94.074509;%%
  %8 citations counted in INSPIRE as of 07 Nov 2016
\bibitem{Aubert:2003fg} 
  B.~Aubert \textit{et al.} [BaBar Collaboration],
  %``Observation of a narrow meson decaying to $D_s^+ \pi^0$ at a mass of 2.32-GeV/c$^2$,''
  Phys.\ Rev.\ Lett.\  \textbf{90}, 242001 (2003).
  %%CITATION = doi:10.1103/PhysRevLett.90.242001;%%
  %752 citations counted in INSPIRE as of 16 Jun 2016
\bibitem{Mohler:2013rwa} 
  D.~Mohler, C.~B.~Lang, L.~Leskovec, S.~Prelovsek and R.~M.~Woloshyn,
  %``$D_{s0}^*(2317)$ Meson and $D$-Meson-Kaon Scattering from Lattice QCD,''
  Phys.\ Rev.\ Lett.\  \textbf{111}, no. 22, 222001 (2013).
  %%CITATION = doi:10.1103/PhysRevLett.111.222001;%%
  %67 citations counted in INSPIRE as of 16 Jun 2016
\bibitem{Lang:2014yfa} 
  C.~B.~Lang, L.~Leskovec, D.~Mohler, S.~Prelovsek and R.~M.~Woloshyn,
  %``Ds mesons with DK and D*K scattering near threshold,''
  Phys.\ Rev.\ D \textbf{90}, no. 3, 034510 (2014).
  %%CITATION = doi:10.1103/PhysRevD.90.034510;%%
  %52 citations counted in INSPIRE as of 16 Jun 2016
\bibitem{Aoki:2008sm}
S.~Aoki \textit{et~al.}, Phys. Rev. D \textbf{79}, 034503 (2009).
\bibitem{Morningstar:2011ka}
C.~Morningstar \textit{et~al.}, Phys. Rev. D \textbf{83}, 114505 (2011).
\bibitem{ElKhadra:1996mp}
A.~X. El-Khadra, A.~S. Kronfeld and P.~B. Mackenzie, Phys. Rev. D \textbf{55}, 3933 (1997).
%%CITATION = HEP-LAT/9604004;%%
%%CITATION = 1104.3870;%%
\bibitem{Luscher} 
  M.~L\"uscher,
  %``Volume Dependence of the Energy Spectrum in Massive Quantum Field Theories. 2. Scattering States,''
  Commun.\ Math.\ Phys.\  \textbf{105}, 153 (1986); Nucl.\ Phys.\ B \textbf{354},
  531 (1991); Nucl.\ Phys.\ B \textbf{364}, 237 (1991).
  %%CITATION = doi:10.1007/BF01211097;%%
  %737 citations counted in INSPIRE as of 20 Jun 2016
\bibitem{Oktay:2008ex} 
  M.~B.~Oktay and A.~S.~Kronfeld,
  %``New lattice action for heavy quarks,''
  Phys.\ Rev.\ D \textbf{78}, 014504 (2008).
  %%CITATION = doi:10.1103/PhysRevD.78.014504;%%
  %62 citations counted in INSPIRE as of 20 Jul 2016
\bibitem{Agashe:2014kda} 
  K.~A.~Olive {\it et al.} [Particle Data Group Collaboration],
  %``Review of Particle Physics,''
  Chin.\ Phys.\ C {\bf 38}, 090001 (2014).
  %%CITATION = doi:10.1088/1674-1137/38/9/090001;%%
  %5118 citations counted in INSPIRE as of 27 Oct 2016
\bibitem{D0:2016mwd} 
  V.~M.~Abazov {\it et al.} [D0 Collaboration],
  %``Evidence for a $B_s^0 \pi^\pm$ state,''
  Phys.\ Rev.\ Lett.\  {\bf 117}, no. 2, 022003 (2016).
  %%CITATION = doi:10.1103/PhysRevLett.117.022003;%%
  %68 citations counted in INSPIRE as of 07 Nov 2016
\bibitem{Aaij:2016iev} 
  R.~Aaij {\it et al.} [LHCb Collaboration],
  %``Search for Structure in the $B_s^0\pi^\pm$ Invariant Mass Spectrum,''
  Phys.\ Rev.\ Lett.\  {\bf 117}, no. 15, 152003 (2016).
  %%CITATION = doi:10.1103/PhysRevLett.117.152003;%%
  %11 citations counted in INSPIRE as of 27 Oct 2016
\end{thebibliography}
\end{document}